\documentclass[conference]{IEEEtran}
\IEEEoverridecommandlockouts
\usepackage{cite}

\usepackage[utf8]{inputenc}
\usepackage{url}
\usepackage{etoolbox}
\usepackage{amsmath,amssymb,amsfonts}
\usepackage{graphicx}
\usepackage{textcomp}
\usepackage{xcolor}
\usepackage{listings}
\usepackage[normalem]{ulem}
 
\def\bibtex{{\rm B\kern-.05em{\sc i\kern-.025em b}\kern-.08em
    T\kern-.1667em\lower.7ex\hbox{E}\kern-.125emX}}
\begin{document}

\title{Empowering Africa: An In-depth Exploration of the Adoption of Artificial Intelligence Across the Continent}

\DeclareRobustCommand*{\IEEEauthorrefmark}[1]{\raisebox{0pt}[0pt][0pt]{\textsuperscript{\footnotesize #1}}}

\author{\IEEEauthorblockN{*Kinyua Gikunda, Denis Kute\\
		Dedan Kimathi University of Technology, Kenya\\
patrick.gikunda@dkut.ac.ke, kute@osanoassociates.com } 
}
\maketitle
\begin{abstract} 
This paper explores the dynamic landscape of Artificial Intelligence (AI) adoption in Africa, analysing its varied applications in addressing socio-economic challenges and fostering development. Examining the African AI ecosystem, the study considers regional nuances, cultural factors, and infrastructural constraints shaping the deployment of AI solutions. Case studies in healthcare, agriculture, finance, and education highlight AI's transformative potential for efficiency, accessibility, and inclusivity. The paper emphasizes indigenous AI innovations and international collaborations contributing to a distinct African AI ecosystem. Ethical considerations, including data privacy and algorithmic bias, are addressed alongside policy frameworks supporting responsible AI implementation. The role of governmental bodies, regulations, and private sector partnerships is explored in creating a conducive AI development environment. Challenges such as digital literacy gaps and job displacement are discussed, with proposed strategies for mitigation. In conclusion, the paper provides a nuanced understanding of AI in Africa, contributing to sustainable development discussions and advocating for an inclusive and ethical AI ecosystem on the continent.
\end{abstract}

\begin{IEEEkeywords}
Artificial Intelligence (AI) Adoption, Socio-economic Challenges, African AI Ecosystem, Ethical Considerations.  
\end{IEEEkeywords}

\section{Introduction} 
The exponential growth of Artificial Intelligence (AI) adoption in Africa heralds a transformative era, where technological innovations are becoming instrumental in tackling pressing socio-economic challenges and driving developmental initiatives \cite{santosh2022artificial}. In 2017, Price Waterhouse Coopers released a report forecasting a substantial contribution of US\$15.7 trillion to the global economy by 2030 \cite{rao2017sizing}. However, the distribution of wealth and influence derived from AI technologies has been notably uneven. PWC's depiction of the future AI-driven economy, as illustrated below, highlights minimal growth for the African continent. Despite this, Africa plays a pivotal role in the development of AI systems, contributing natural resources, labor, and skills from across the region. Despite the extensive global reach of the AI supply chain, the advantages of these technologies have yet to materialize in Africa. Instead, they predominantly benefit Big Tech companies in the Global North and China, along with those who can afford the daily conveniences offered by AI through devices like Amazon's Alexa and smart cars.

The African continent presents a distinctive context poised to harness AI technologies for local socio-economic advancement \cite{ndubisi2022artificial}. With the potential to leapfrog the technological infrastructure associated with the Third Industrial Revolution and a dynamic youth population adept at adapting to digital work and entrepreneurship, Africa is well-positioned for growth. Emphasizing the importance of prioritizing AI solutions aligned with national developmental priorities, African governments should strive for prosperous and inclusive societies.

While policy responses to AI are emerging in Africa, illustrated by countries like Egypt, Mauritius, and Rwanda publishing national AI strategies, the continent remains predominantly influenced by foreign technology and AI firms \cite{adams2022ai}. These external entities may not necessarily align with national developmental goals, and in some cases, may exacerbate exclusion and oppression, particularly affecting women. To counter this, African policymakers must prioritize the development of local AI capabilities that can significantly contribute to inclusive economic growth and social transformation. This emphasis necessitates building AI policy responses on national digital agendas and ensuring equitable access to digital, data, and computing infrastructure.

In an analysis carried out between two countries —one advanced and the other developing- to find out if AI has the potential to widen the gap between rich and poor nations, the authors identify three significant channels through which divergence between the two economies can occur \cite{santosh2022artificial}. Firstly, in terms of share-in-production, advanced economies with higher total factor productivity and wages use AI more intensively, leading to a greater long-term benefit when AI productivity increases. This divergence is more pronounced when AI easily substitute for workers. Secondly, investment flows are impacted as the productivity increase in with use of AI creating higher demand for investment in both robots and traditional capital, particularly in advanced economies. This results in a diversion of investment from developing countries, causing a transitional decline in their GDP. Lastly, the terms-of-trade channel suggests that a developing economy, specializing in sectors reliant on unskilled labor, may experience a permanent decline in terms of trade after the AI revolution. As AI replace unskilled labor, leading to a drop in their relative wages, the price of goods using unskilled labor intensively decreases, potentially causing a negative impact on investment incentives and resulting in a fall in both relative and absolute GDP. These factors collectively illustrate the multifaceted challenges faced by developing countries in the wake of the AI revolution.

In a broader context, the brief adopts a comprehensive definition of AI to encompass the entire AI life cycle, accounting for both the potential harms and benefits. This life cycle spans from the raw materials and labor required to build the infrastructure and data systems supporting AI production to the design, development, rollout, and review of AI technologies in societies. The paper conducts a comprehensive analysis of the multifaceted impact of AI integration in Africa, offering a holistic examination of the diverse applications shaping the continent's developmental landscape. The study's commitment to comprehensiveness is evident through its in-depth exploration of how AI technologies are strategically employed, moving beyond mere adoption to address socio-economic challenges. By delving into real-world applications and case studies, the paper provides tangible examples of AI contributions in key sectors crucial to Africa's development. The exploration recognizes the unique characteristics of the African AI ecosystem, emphasizing regional variations, cultural nuances, and infrastructural constraints. This nuanced understanding goes beyond a one-size-fits-all approach, acknowledging the pivotal role of regional disparities in shaping AI efficacy. The consideration of cultural nuances underscores the socio-cultural dimension of AI adoption, stressing the importance of aligning technological solutions with diverse cultural fabrics. Additionally, the paper highlights the significance of infrastructural investments for the successful deployment and impact of AI in Africa. In essence, this commendable exploration navigates the intersection of technology, society, and development, offering a foundational understanding crucial for navigating the evolving dynamics of Africa's technological journey.

The paper is structured as follows: section \ref{Emerging AI Technologies Impact and Concerns in Africa} we present Emerging AI Technologies Impact and Concerns in Africa, section \ref{Existing AI Policy Landscape in Africa} discussion on Existing AI Policy Landscape in Africa , section \ref{AI-related Policy Areas in Africa} AI-related Policy Areas in Africa, section \ref{Strategic Planning for AI in Africa} Strategic Planning for AI in Africa and section \ref{Conclusion} conclusion.

\section{Emerging AI Technologies Impact and Concerns in Africa}
\label{Emerging AI Technologies Impact and Concerns in Africa}
AI has the potential to fundamentally change the way businesses operate, drive innovation, and improve the lives of millions of people across Africa \cite{wamba2020influence}. AI is set to be the key source of transformation, disruption and competitive advantage in today’s fast changing economy. Some of the key sectors that could benefit from AI include healthcare \cite{owoyemi2020artificial}, agriculture \cite{aguera2020paving}, water \cite{valipour2015future}, education \cite{pedro2019artificial}, and finance \cite{mhlanga2020industry}. There are already a number of applications of AI in Africa, especially towards smart farming \cite{foster2023smart}, health \cite{shaheen2021applications}, water supply \cite{macharia2020applying}, clean energy forecasting \cite{motepe2019improving}, climate change predictions \cite{rutenberg2021use}, finance \cite{mhlanga2021financial} and governance \cite{plantinga2022digital}. AI-powered products and systems are ubiquitous throughout the continent. Locally created AI systems are being created in Kenya to help farmers with equipment purchases and decision-making around the best time to plant or harvest particular crops \cite{foster2023smart}. Like many other African cities, Lagos, Nigeria, is seeing a rise in the number of young data scientists who come together to create new AI-driven companies and technology and hone their machine learning skills. However, AI technologies pose potential opportunities and threats to the society. This section presents areas with the biggest potential of AI in Africa along with some of the challenges faced to realise the full potential of its implementation. 

AI-powered diagnostics use the patient's individual medical history as a baseline; minor variations from this indicate the possibility of a health issue that needs to be looked into and treated \cite{simon2019applying}. AI could be used to support human doctors rather than to replace them. Not only will it improve doctors' diagnosis, but it will also yield insightful data that will help the AI learn and grow over time. Human doctors and AI-powered diagnostics will work together continuously to improve system accuracy, which will eventually give people enough confidence to give the duty to the AI system so it can run on its own \cite{ahuja2019impact}. Areas with greatest AI potential include supporting diagnosis such as comparing patient data with similar patients or identifying little deviations from the baseline in patients' health data. Preventing and controlling the spread of diseases through early detection of possible pandemics and monitoring illness incidence. The primary advantages are quicker and more precise diagnosis and more individualized care in the short- and medium-term, which would open the door for longer-term advancements in fields like intelligent implants. Addressing concerns related to the privacy and safeguarding of sensitive health data is imperative. The intricacies of human biology and the necessity for continued technological advancements imply that the realization of full potential and acceptance from patients, healthcare providers, and regulators for some of the more advanced applications may require time.

AI advancements like robo-advice have made it feasible to create customized investment solutions for mass market customers in ways that were previously only accessible to high net worth clients, even though human financial guidance is expensive and time-consuming \cite{shanmuganathan2020behavioural}. Finances are managed dynamically to maximize clients' available finances and align with goals (such saving for a mortgage), as asset managers are supplemented and, in certain situations, completely replaced by AI. The data and technology are in place, but to reach its full potential, client acceptability must yet rise. The realm of AI in financial services encompasses a spectrum of applications aimed at enhancing efficiency and security. Personalized financial planning leverages AI algorithms to tailor financial advice and services to individual needs, optimizing investment strategies and financial decision-making. Additionally, AI plays a pivotal role in fraud detection and anti-money laundering efforts, employing advanced algorithms to analyze transaction patterns and identify potential irregularities, thereby fortifying the security of financial systems. Beyond the back-office functions, AI facilitates process automation in customer-facing operations, streamlining interactions and improving the overall customer experience. This multifaceted integration of AI in financial services reflects a transformative approach, where technology is not only enhancing operational efficiency but also customizing financial solutions for individual users while ensuring the integrity of the financial ecosystem. Challenges in AI to Address includes establishing Consumer trust and gaining regulatory acceptance

Three key areas stand out as having significant potential for AI in retail industry \cite{oosthuizen2021artificial}. First, personalized design and production hold promise for a future where products can be tailored to individual preferences and needs. Retailers are increasingly leveraging deep learning to anticipate customer demand, predicting orders in advance. This not only streamlines inventory and delivery management but also aligns with the consumer benefit of on-demand customization, offering greater availability of desired products. In terms of timing, product recommendations based on preferences are already a reality, while the medium-term potential includes fully customized products. In the longer term, AI-driven products may anticipate market demand signals. The time saved for consumers is notable, as AI eliminates the need to extensively explore shelves, catalogues, or websites. However, barriers to overcome include adapting design and production to this agile and tailored approach, along with the imperative for businesses to bolster trust regarding data usage and protection. A high-potential use case is personalized design and production, particularly in sectors like fashion, where AI could facilitate interactive and customized design and supply processes based on user feedback.

Three major areas demonstrate significant AI potential in technology and entertainment industry: media archiving and search, customised content creation, and personalized marketing and advertising \cite{chan2019review}. In media archiving and search, AI is instrumental in efficiently categorizing and recommending content from vast and diverse sources. Customised content creation spans various domains like marketing, film, and music, allowing for tailored content generation. Personalized marketing and advertising leverage AI to cater to individual preferences, offering a more targeted approach. Consumers benefit from increasingly personalized content generation and recommendations \cite{anantrasirichai2022artificial}. In terms of timing, content recommendation is already prevalent, while the medium-term potential includes automated telemarketing capable of holding real conversations with customers. The longer-term vision involves AI creating use-case specific and individualized content. Time savings for consumers stem from quicker and easier content selection aligned with their preferences and mood. However, challenges include navigating through the abundance of unstructured data. A high-potential use case is seen in media archiving and search, where AI streamlines content classification, facilitating precise targeting and increased revenue generation.

Three primary areas highlight substantial AI potential in manufacturing: enhanced monitoring and auto-correction of manufacturing processes, supply chain and production optimization, and on-demand production \cite{arinez2020artificial}. The consumer benefits from a more flexible, responsive, and custom-made manufacturing of goods, leading to fewer delays, defects, and faster delivery. Currently, greater automation of numerous production processes is readily achievable. In the medium term, intelligent automation can extend to areas like supply chain optimization and predictive scheduling. The longer-term vision involves using prescriptive analytics in product design, going beyond predicting and responding to demand, to actively solve problems and shape outcomes. AI's role in facilitating seamless integration of supply chain data enables anticipatory production and more efficient product delivery, resulting in time savings and faster responses. However, challenges include the need for all parties to possess the necessary technology and collaborate effectively, with currently only the largest and best-resourced suppliers and manufacturers being at the forefront. A high-potential use case lies in enhanced monitoring and auto-correction, where self-learning systems enhance predictability and control in the manufacturing process, reducing delays, defects, and deviations from specifications.

Identifying three key areas with significant AI potential in the energy sector includes smart metering, more efficient grid operation and storage, and predictive infrastructure maintenance \cite{ahmad2021artificial}. The consumer benefits from these applications through a more efficient and cost-effective supply and usage of energy. Smart metering, providing real-time information on energy usage, is already implementable. In the medium term, potential lies in optimized power management, while the longer-term vision includes more efficient and consistent renewable energy supply, particularly in the improved prediction and optimization of wind power. The time saved for consumers results from a more secure energy supply and fewer outages. However, challenges include technological development and high investment requirements in some of the more advanced areas. A high-potential use case is observed in smart meters, as they enable customers to tailor their energy consumption, reduce costs, and contribute to a wealth of data that could pave the way for more customized tariffs and a more efficient energy supply.

In the realm of transport and logistics, three primary areas exhibit significant AI potential: autonomous trucking and delivery, traffic control and reduced congestion, and enhanced security \cite{khang2023ai}. Consumers stand to benefit from greater flexibility, customization, and choice in the movement of goods and people, leading to faster and more reliable transportation. While automated picking in warehouses is already implementable, the medium-term potential lies in traffic control, and the longer-term vision involves autonomous trucking and delivery. Time saved results from smart scheduling, reduced traffic jams, and real-time route adjustments to expedite transport. However, challenges persist, particularly in the development of technology for autonomous fleets. A high-potential use case is observed in traffic control and reduced congestion, where autonomous trucking not only cuts costs through increased asset utilization but also disrupts the traditional transport and logistics business model. The application of augmented intelligence is exemplified by an automotive company using a dynamic agent-based model to simulate strategic scenarios, providing decision-makers with a risk-free environment to understand the impact of various policies on market share and revenue. This approach parallels the use of flight simulators for pilots, emphasizing the importance of simulated environments for executive decision-making.

\section{Existing AI Policy Landscape in Africa}
\label{Existing AI Policy Landscape in Africa}
Countries across various regions of Africa are increasingly formulating or considering national AI strategies to guide the adoption of AI. These strategies align with global trends in AI policies \cite{ravizki2023legal}. These trends encompass diverse aspects, including basic and applied research in AI, talent attraction, development, and retention, considerations for the future of work and skills, the industrialization of AI technologies, public sector utilization of AI, data and digital infrastructure, ethical considerations, regulatory frameworks, inclusion, and foreign policy. Several African nations are incorporating AI-relevant policies within digital or data policy frameworks. The paper offers an overview of these emerging trends, with a focus on national AI strategies in Mauritius, Egypt, and Rwanda, as well as the Blueprint for AI in Africa developed by the pan-African government-endorsed group SMARTAfrica. The discussion extends to explore evolving AI-related policy mechanisms in other African countries.

Mauritius emerged as the trailblazer among African nations by publishing the first AI strategy document in 2018, titled the Mauritius AI Strategy \cite{mauritiusai}. Produced by the national Working Group on AI, the strategy envisions AI as a catalyst for creating a new developmental pillar for the nation, driving socio-economic growth across key sectors. The document outlines priority AI areas, with a specific focus on manufacturing, healthcare, fintech, and agriculture. Additionally, it highlights the potential applications and scoping of existing projects within these domains. The strategy extends its focus to AI applications in the Ocean Economy, emphasizing smart ports and traffic management, as well as in energy management to contribute to CO2 emissions reduction. The report concludes with a set of recommendations covering crucial aspects such as skills development, research and development funding, and governance mechanisms, including provisions for data protection, open data platforms, and the establishment of an AI ethics committee.

In July 2021, Egypt unveiled a comprehensive National AI Strategy to be implemented over the next three to five years \cite{egypt_2023}. The central objective of the strategy is to ensure that AI adoption aligns with Egypt's national developmental priorities, focusing on enhancing government efficiencies, transparency, and decision-making. However, there remains a question regarding whether the strategy adequately addresses human rights concerns related to mass government surveillance and censorship. The strategy is built upon four pillars: advancing AI use in government, promoting AI adoption by the public and private sector for development goals, building AI capacities through education and research, and taking a leading role in international and regional AI cooperation. These pillars are supported by four enablers: governance, including laws, policies, and ethical guidance; data, incorporating the Personal Data Protection Law and a forthcoming data strategy; infrastructure, encompassing cloud computing and data storage; and ecosystems, fostering institutions, startups, and talent for a responsible and innovative national AI ecosystem.

The National AI Policy for the Republic of Rwanda serves as a strategic road-map to leverage the advantages of AI while addressing associated risks \cite{rwanda_2023}. Aligned with the Vision 2050, Smart Rwanda Master Plan, and other national plans, the policy aims to empower Rwanda in utilizing AI for sustainable and inclusive growth. Through collaboration with local, regional, and international stakeholders, the policy aspires to position Rwanda as a leading African Innovation Hub and the continent's Center of Excellence in AI. The policy consist of six fundamental elements. Firstly, it concentrates on establishing the groundwork for AI adoption by cultivating 21st-century skills and AI literacy, ensuring robust infrastructure and compute capacity, and formulating a comprehensive data strategy. Secondly, the policy underscores the significance of responsible AI adoption, emphasizing fairness, transparency, and accountability in AI systems. Thirdly, it acknowledges the pivotal role of the private sector in AI adoption, promoting collaboration between the government and private entities to drive AI innovation. Fourthly, the policy advocates for Rwanda's active participation in international discussions on AI governance and ethics, recognizing AI as a global issue. The fifth element involves monitoring and evaluation, incorporating mechanisms to assess policy implementation and ensure effectiveness in achieving set objectives. Lastly, the policy is designed as a dynamic document, subject to periodic review and revision to stay abreast of evolving AI technology.

SmartAfrica\footnote{https://smartafrica.org/}, established in 2013 by African heads of state, aims to advance information and communication technologies and digital development in Africa. In collaboration with the South African government, SmartAfrica published the Blueprint on AI for Africa in 2021. This Blueprint identifies AI as a technology influencing all aspects of society and the economy, outlining five pillars for a successful AI strategy in African countries. These pillars include human capital development, the transition of AI solutions from lab to market, essential infrastructure for local AI development, networking through expanded ecosystems, and regulations to address emerging AI challenges and opportunities at national and sectoral levels. The Blueprint also highlights key AI use cases across various sectors. The adoption of such AI policy frameworks in Africa addresses concerns related to data protection, emphasizes building national AI ecosystems, and seeks to serve local needs by countering foreign tech influences. While there is emphasis on protecting labor markets from wholesale AI-driven automation, attention to the labor rights of digital workers and potential AI-related threats to democracy and human rights remains insufficient. The subsequent discussion delves into developing AI-related policy areas in Africa, evaluating their potential to address emergent issues and considering the opportunities and limitations of seeking guidance from the EU on AI governance frameworks.

\section{AI-related Policy Areas in Africa}
\label{AI-related Policy Areas in Africa}
\subsection{Data Governance and Protection}
Access to data is a pivotal element in the development of AI in Africa \cite{amankwah2022harnessing}. To foster the creation of AI solutions addressing local developmental challenges, AI developers in Sub-Saharan Africa need extensive access to relevant and high-quality data. However, the Open Data Barometer\footnote{https://opendatabarometer.org} indicates that the region lags behind others in providing meaningful access to public sector data \cite{lemma2022afcfta}. The data-focused policies in the AI strategies of Egypt and Rwanda, along with the adoption of the Continental Data Policy Framework by the African Union Commission, are commendable steps toward ensuring equitable access to digital and data-driven technologies for all Africans.

Data protection laws have a significant impact on attitudes toward sharing data containing personal information \cite{brand2022data}. The European General Data Protection Regulation\footnote{https://gdpr-info.eu/} imposes strict adequacy requirements on sharing personal data outside the European Union. Simultaneously, an emerging trend in Africa sees countries establishing data protection laws or policies to retain data within their territories \cite{Babalola2023}. This trend is driven by economic considerations, recognizing data as a crucial innovation resource that fuels market efficiencies and economic growth. Keeping data within the country facilitates easier access by local groups and responds to concerns that digital companies operating in the region may bring limited local economic benefits without establishing a physical presence.

For instance, the Kenyan Data Protection Act of 2019\footnote{https://www.odpc.go.ke/dpa-act/} stipulates that a copy of all personal data transferred outside the country must also be stored within Kenya. The South African Draft Data and Cloud Policy goes further, requiring all data generated in South Africa to be stored within the country \cite{shibambu2022framework}. However, such policies, including those in Nigeria, promoting data localization, have faced criticism. Critics argue that these policies could legitimize mass government surveillance and run counter to best practices related to the stewardship, rather than ownership, of data \cite{kugler2022impact}.

\subsection{Re-skilling}
In the rapidly evolving landscape of AI, reskilling has emerged as a critical policy area in Africa. As the continent embraces the transformative potential of AI technologies, there is a growing recognition that the existing workforce needs to be equipped with new skills to navigate the changing employment landscape \cite{mukherjee2023edge}. Reskilling initiatives are vital to address the potential challenges posed by automation, ensuring that individuals are not left behind as job roles evolve with the integration of AI. Policymakers are actively engaged in crafting strategies that promote continuous learning, offering training programs and educational resources to empower workers with the expertise needed to collaborate effectively with AI systems.

Reskilling policies in Africa should extend beyond individual skill development to encompass broader societal goals. These initiatives align with the vision of creating inclusive and resilient economies, where the benefits of AI are distributed equitably. By prioritizing reskilling, policymakers aim to foster a workforce that is not only adaptable to technological advancements but also capable of driving innovation and leveraging AI for sustainable economic growth. The emphasis on reskilling reflects a proactive approach to harnessing the opportunities presented by AI, ensuring that Africa's human capital remains at the forefront of the global digital transformation. 

Traditional policy responses to the threat of job loss from automation often do not align well with many African contexts \cite{marwala2022closing}. The prevailing approach involves implementing programs to re-skill or up-skill existing workforces, but this response does not adequately address the impact of automation on job loss for African women. Women, who predominantly hold low-skilled labor and repetitive tasks, face a higher risk of being replaced by automation. Unfortunately, the experiences of women are frequently overlooked in policy measures, particularly in reskilling programs that fail to consider the daily realities of women who bear domestic responsibilities and have limited time to reskill for the digital world. An illustrative case is South Africa's 2019 White Paper on Science, Technology, and Innovation, which does not explicitly address the specific challenges faced by women in the context of job loss due to automation \cite{beumer2022economic}. These challenges are compounded by the growing digital gender divide within the continent. Social factors, including deep-seated patriarchy, contribute to women having poorer access to digital technologies compared to men. 

\subsection{Ethical Guidelines}
Ethical guidelines in the context of AI play a pivotal role in ensuring responsible and humane development, deployment, and use of AI technologies \cite{roche2023ethics}. As AI systems become increasingly sophisticated and integrated into various aspects of society, ethical considerations become paramount. Ethical guidelines are designed to set standards and principles that guide developers, policymakers, and organizations in creating AI systems that align with values such as fairness, transparency, accountability, and inclusivity. These guidelines address the potential biases embedded in AI algorithms, emphasizing the need to mitigate discriminatory impacts and ensure equitable outcomes. Additionally, they underscore the importance of transparency in AI decision-making processes, allowing users and stakeholders to understand how AI systems arrive at specific conclusions or recommendations.

A robust set of ethical guidelines also considers accountability mechanisms, establishing frameworks for responsible AI development and use. Developers and organizations are encouraged to take responsibility for the consequences of their AI systems, including addressing any unintended negative impacts on individuals or communities. Inclusivity is another key dimension, urging developers to design AI systems that consider diverse perspectives and avoid reinforcing existing societal biases. Ethical guidelines are dynamic documents that evolve with technological advancements and emerging ethical challenges. They serve as a foundation for fostering public trust in AI technologies and are instrumental in shaping a future where AI benefits humanity while respecting fundamental ethical principles.

\subsection{Social Media Regulation}
The intersection of social media and AI introduces a unique set of policy considerations. As social media platforms increasingly utilize AI algorithms for content moderation, recommendation systems, and user interactions, African nations are grappling with the need for comprehensive policies to address the challenges and opportunities presented by this technological convergence \cite{plantinga2022digital}. 

One key policy area is the regulation of content and misinformation on social media platforms. AI algorithms play a crucial role in content moderation, identifying and removing inappropriate or harmful content. However, the effectiveness of these algorithms raises concerns about potential biases and limitations, necessitating policies that ensure fairness, transparency, and accountability. African governments should explore ways to develop AI-related content regulations that align with local cultural contexts and societal norms, avoiding the imposition of global standards that may not be suitable for diverse African communities \cite{plantinga2023responsible}.

Another vital aspect is data privacy and protection in the context of AI-driven social media platforms. As AI relies heavily on vast amounts of user data, there is a growing emphasis on implementing robust data protection regulations \cite{prinsloo2022data}. African nations are formulating policies to safeguard user privacy, determine the permissible uses of personal data, and regulate data sharing practices between social media companies and third parties. Ensuring that AI-driven social media platforms prioritize data security and user privacy is crucial for building trust among African users.

\section{Strategic Planning for AI in Africa}
\label{Strategic Planning for AI in Africa}
As African policymakers navigate the landscape of AI strategies and adoptions, the focal point should be on establishing sustainable local AI ecosystems that actively contribute AI solutions to advance national developmental priorities, fostering inclusive and prosperous African societies. This strategic approach transcends the economic dimensions of AI and requires a critical examination of how global AI frameworks address the unique challenges posed in the region. Building effective policy responses necessitates the involvement of a diverse range of stakeholders, including local tech and data entrepreneurs, as well as social justice groups and communities directly affected by AI policies. By adopting this comprehensive and inclusive approach, African nations can tailor their AI strategies to align with local needs and priorities, ensuring that the benefits of AI adoption are harnessed for the overall well-being and development of their societies.
\subsection{Digital Infrastructure Development}
Policies related to data governance, including data protection laws, are crucial for creating a framework that ensures responsible and ethical use of data in AI applications. Efforts to keep data within national borders and comply with international data protection standards are emerging trends. African governments should prioritize policy efforts directed at establishing and sustaining safe, secure, and inclusive infrastructure to facilitate the local development of AI. This entails policies aimed at enhancing internet access and preventing local entities from implementing internet shutdowns. Additionally, policies supporting good governance and the availability of data for development are crucial. The potential adoption of the African Union Data Policy Framework holds significance in promoting uniform data governance standards continent-wide. The Africa-EU Global Gateway serves as a vital mechanism for ensuring equitable access to both basic and advanced digital infrastructure, underscoring the importance of leveraging this platform for collaborative efforts in the development and maintenance of digital resources.
\subsection{Regional Co-Operation}
Collaborative efforts at the regional level are critical for developing common regulatory responses, taxation provisions for multinational tech companies, and data-sharing agreements between countries. The African Continental Free Trade Area Protocol on E-Commerce is one example. Emphasizing the importance of regional cooperation, it stands as a vital policy approach for creating unified regulatory measures applicable to multinational and foreign tech companies operating within the region. This collaboration can extend to the development of taxation provisions tailored for multinational social media platforms. Moreover, regional cooperation is crucial for establishing data-sharing agreements between countries, enabling access to a broader spectrum of public sector data. This shared data can prove instrumental for local AI developers in advancing developmental priorities. The African Continental Free Trade Area Protocol on E-Commerce represents a significant opportunity to integrate provisions supporting interregional data sharing, thereby contributing to regional development goals and fostering economic growth.

\subsection{Building Local Capacity and Skills}
The imperative development of AI, along with the necessary data and technology skills among policy-makers and the workforce in Africa, is foundational for fostering the responsible use and advancement of AI. Comprehensive capacity development policies are essential to cultivate a thorough understanding of AI at all levels. Specific policy measures should be implemented to promote gender inclusivity, particularly advancing women in STEM (Science, Technology, Engineering, and Mathematics) fields and AI-related decision-making positions. Furthermore, policies in this domain may encompass measures to attract diverse AI talent by removing entry barriers for Africans possessing skills in data science and computing. This approach aims to ensure a skilled and inclusive workforce capable of harnessing the benefits of AI for the continent's development.

Labor market policies in AI-related policy areas across Africa are gaining traction as nations grapple with the transformative impact of AI. Recognizing the potential disruption and opportunities posed by AI, African countries are prioritizing policies aimed at addressing the challenges in the labor sector. These policies encompass reskilling and upskilling programs to equip the workforce with the necessary skills for an AI-driven economy. Efforts are also directed towards ensuring inclusivity, with a particular focus on addressing the unique challenges faced by women in low-skilled roles susceptible to automation. Moreover, the policies extend beyond individual skill development to encompass the protection of digital workers' rights, navigating the complexities of gig economies and click-labor. The overarching goal is to establish a resilient and adaptive labor market that maximizes the benefits of AI while safeguarding the well-being and rights of workers across diverse sectors in Africa.

\subsection{Public Awareness, Participation and Beneficiation}
Policies aimed at fostering public awareness, understanding, and involvement in AI-related discussions and decisions contribute to creating a more inclusive and transparent policy environment. Given the substantial economic potential of AI and the broad societal implications associated with certain AI applications, it is crucial to actively engage local communities in decisions concerning the design and implementation of AI systems that could impact them. Additionally, the formulation of beneficiation frameworks becomes imperative to ensure that communities whose data contribute to the development of an AI system receive tangible benefits. In this context, beneficiation should be a foundational principle embedded in national AI ethical guidelines, emphasizing the importance of ensuring that the advantages of AI are shared equitably among the communities involved in its development and application.

\subsection{International Development Assistance}
International development assistance plays a pivotal role in shaping AI-related policy areas in Africa, offering a valuable opportunity for collaborative efforts to ensure responsible and inclusive AI adoption across the continent. International development assistance remains a crucial avenue for fostering African development. To facilitate the adoption of responsible AI solutions across the continent, donors, intergovernmental organizations, and other funding entities should focus their efforts on aiding initiatives that promote inclusive digital infrastructure and cultivate sustainable local expertise in AI governance. An illustrative example is the Africa-EU Global Gateway investment scheme, which exemplifies the commitment to these objectives. Additionally, it is essential to underscore the importance of preserving the sovereignty of African states in crafting AI governance frameworks. Such frameworks should align with national constitutional principles and foundational values, actively contributing to the realization of local development priorities.

A key emphasis within the realm of international development assistance is the promotion of responsible AI solutions. Donors recognize the importance of ensuring that AI technologies align with ethical principles, human rights, and local values. Consequently, funding and support are directed towards initiatives that prioritize the development of responsible AI frameworks, guidelines, and governance structures. This includes capacity-building programs to enhance the understanding of AI among policymakers and communities, fostering an environment where AI is utilized for positive societal impact. The Africa-EU Global Gateway investment scheme stands as a notable example of international collaboration in supporting AI-related policies in Africa. This initiative reflects a commitment to fostering equal access to both basic and advanced digital infrastructure across the continent. By investing in digital connectivity and technology capabilities, international development assistance aims to bridge the digital divide, ensuring that African nations can actively participate in the global digital economy and fully realize the potential benefits of AI technologies.

Crucially, international development assistance also underscores the importance of respecting the sovereignty of African states in developing AI governance solutions. Policies and frameworks are encouraged to align with national constitutional principles and address local development priorities. This approach ensures that AI-related initiatives are not only technologically advanced but also culturally sensitive, fostering a collaborative and inclusive approach to AI adoption in Africa.

\section{Conclusion} 
\label{Conclusion} 
In conclusion, the exploration of the emerging applications of AI across the African continent underscores both the vast potential and unique challenges that this transformative technology presents. As African nations strategically adopt AI, focusing on areas like smart metering, personalized marketing, and infrastructure optimization, there is a palpable opportunity for positive socio-economic growth and inclusive development. However, it is crucial for policymakers and stakeholders to navigate this landscape with a nuanced understanding of regional variations, cultural nuances, and infrastructural constraints. The development of national AI strategies, exemplified by countries like Mauritius, Egypt, and Rwanda, marks a pivotal step, yet challenges such as data access, protection, and local capacity-building remain paramount. The collective effort to build sustainable local AI ecosystems, ensure ethical considerations, and prioritize the inclusivity of marginalized groups is essential. By fostering regional cooperation, promoting capacity development, and engaging local communities, Africa can harness the benefits of AI while safeguarding its sovereignty and advancing national development priorities. International collaboration, inclusive digital infrastructure, and responsible governance practices are pivotal elements in shaping a future where AI contributes to a prosperous and inclusive Africa.
\bibliographystyle{IEEEtran}
\bibliography{bibtext}

\begin{thebibliography}{10}
\providecommand{\url}[1]{#1}
\csname url@samestyle\endcsname
\providecommand{\newblock}{\relax}
\providecommand{\bibinfo}[2]{#2}
\providecommand{\BIBentrySTDinterwordspacing}{\spaceskip=0pt\relax}
\providecommand{\BIBentryALTinterwordstretchfactor}{4}
\providecommand{\BIBentryALTinterwordspacing}{\spaceskip=\fontdimen2\font plus
\BIBentryALTinterwordstretchfactor\fontdimen3\font minus \fontdimen4\font\relax}
\providecommand{\BIBforeignlanguage}[2]{{%
\expandafter\ifx\csname l@#1\endcsname\relax
\typeout{** WARNING: IEEEtran.bst: No hyphenation pattern has been}%
\typeout{** loaded for the language `#1'. Using the pattern for}%
\typeout{** the default language instead.}%
\else
\language=\csname l@#1\endcsname
\fi
#2}}
\providecommand{\BIBdecl}{\relax}
\BIBdecl

\bibitem{santosh2022artificial}
K.~Santosh and L.~Gaur, \emph{Artificial intelligence and machine learning in public healthcare: Opportunities and societal impact}.\hskip 1em plus 0.5em minus 0.4em\relax Springer Nature, 2022.

\bibitem{rao2017sizing}
A.~S. Rao and G.~Verweij, ``Sizing the prize: What’s the real value of ai for your business and how can you capitalise?'' 2017.

\bibitem{ndubisi2022artificial}
E.~J. NDUBISI and K.~Ikechukwu~Anthony, ``Artificial intelligence and socio-economic development in africa,'' \emph{OCHENDO: An African Journal of Innovative Studies}, vol.~3, no.~1, 2022.

\bibitem{adams2022ai}
R.~Adams, ``Ai in africa: key concerns and policy considerations for the future of the continent,'' 2022.

\bibitem{wamba2020influence}
S.-L. Wamba-Taguimdje, S.~Fosso~Wamba, J.~R. Kala~Kamdjoug, and C.~E. Tchatchouang~Wanko, ``Influence of artificial intelligence (ai) on firm performance: the business value of ai-based transformation projects,'' \emph{Business Process Management Journal}, vol.~26, no.~7, pp. 1893--1924, 2020.

\bibitem{owoyemi2020artificial}
A.~Owoyemi, J.~Owoyemi, A.~Osiyemi, and A.~Boyd, ``Artificial intelligence for healthcare in africa,'' \emph{Frontiers in Digital Health}, vol.~2, p.~6, 2020.

\bibitem{aguera2020paving}
P.~Aguera, N.~Berglund, T.~Chinembiri, A.~Comninos, A.~Gillwald, and N.~Govan-Vassen, ``Paving the way towards digitalising agriculture in south africa,'' \emph{no. June}, pp. 1--42, 2020.

\bibitem{valipour2015future}
M.~Valipour, ``Future of agricultural water management in africa,'' \emph{Archives of Agronomy and Soil Science}, vol.~61, no.~7, pp. 907--927, 2015.

\bibitem{pedro2019artificial}
F.~Pedro, M.~Subosa, A.~Rivas, and P.~Valverde, ``Artificial intelligence in education: Challenges and opportunities for sustainable development,'' 2019.

\bibitem{mhlanga2020industry}
D.~Mhlanga, ``Industry 4.0 in finance: the impact of artificial intelligence (ai) on digital financial inclusion,'' \emph{International Journal of Financial Studies}, vol.~8, no.~3, p.~45, 2020.

\bibitem{foster2023smart}
L.~Foster, K.~Szilagyi, A.~Wairegi, C.~Oguamanam, and J.~de~Beer, ``Smart farming and artificial intelligence in east africa: Addressing indigeneity, plants, and gender,'' \emph{Smart Agricultural Technology}, vol.~3, p. 100132, 2023.

\bibitem{shaheen2021applications}
M.~Y. Shaheen, ``Applications of artificial intelligence (ai) in healthcare: A review,'' \emph{ScienceOpen Preprints}, 2021.

\bibitem{macharia2020applying}
P.~Macharia, N.~Kreuzinger, and N.~Kitaka, ``Applying the water-energy nexus for water supply—a diagnostic review on energy use for water provision in africa,'' \emph{Water}, vol.~12, no.~9, p. 2560, 2020.

\bibitem{motepe2019improving}
S.~Motepe, A.~N. Hasan, and R.~Stopforth, ``Improving load forecasting process for a power distribution network using hybrid ai and deep learning algorithms,'' \emph{IEEE Access}, vol.~7, pp. 82\,584--82\,598, 2019.

\bibitem{rutenberg2021use}
I.~Rutenberg, A.~Gwagwa, and M.~Omino, ``Use and impact of artificial intelligence on climate change adaptation in africa,'' in \emph{African Handbook of Climate Change Adaptation}.\hskip 1em plus 0.5em minus 0.4em\relax Springer, 2021, pp. 1107--1126.

\bibitem{mhlanga2021financial}
D.~Mhlanga, ``Financial inclusion in emerging economies: The application of machine learning and artificial intelligence in credit risk assessment,'' \emph{International journal of financial studies}, vol.~9, no.~3, p.~39, 2021.

\bibitem{plantinga2022digital}
P.~Plantinga, ``Digital discretion and public administration in africa: Implications for the use of artificial intelligence,'' \emph{Information Development}, p. 02666669221117526, 2022.

\bibitem{simon2019applying}
G.~Simon, C.~D. DiNardo, K.~Takahashi, T.~Cascone, C.~Powers, R.~Stevens, J.~Allen, M.~B. Antonoff, D.~Gomez, P.~Keane \emph{et~al.}, ``Applying artificial intelligence to address the knowledge gaps in cancer care,'' \emph{The oncologist}, vol.~24, no.~6, pp. 772--782, 2019.

\bibitem{ahuja2019impact}
A.~S. Ahuja, ``The impact of artificial intelligence in medicine on the future role of the physician,'' \emph{PeerJ}, vol.~7, p. e7702, 2019.

\bibitem{shanmuganathan2020behavioural}
M.~Shanmuganathan, ``Behavioural finance in an era of artificial intelligence: Longitudinal case study of robo-advisors in investment decisions,'' \emph{Journal of Behavioral and Experimental Finance}, vol.~27, p. 100297, 2020.

\bibitem{oosthuizen2021artificial}
K.~Oosthuizen, E.~Botha, J.~Robertson, and M.~Montecchi, ``Artificial intelligence in retail: The ai-enabled value chain,'' \emph{Australasian Marketing Journal}, vol.~29, no.~3, pp. 264--273, 2021.

\bibitem{chan2019review}
S.~M. Chan-Olmsted, ``A review of artificial intelligence adoptions in the media industry,'' \emph{International Journal on Media Management}, vol.~21, no. 3-4, pp. 193--215, 2019.

\bibitem{anantrasirichai2022artificial}
N.~Anantrasirichai and D.~Bull, ``Artificial intelligence in the creative industries: a review,'' \emph{Artificial intelligence review}, pp. 1--68, 2022.

\bibitem{arinez2020artificial}
J.~F. Arinez, Q.~Chang, R.~X. Gao, C.~Xu, and J.~Zhang, ``Artificial intelligence in advanced manufacturing: Current status and future outlook,'' \emph{Journal of Manufacturing Science and Engineering}, vol. 142, no.~11, p. 110804, 2020.

\bibitem{ahmad2021artificial}
T.~Ahmad, D.~Zhang, C.~Huang, H.~Zhang, N.~Dai, Y.~Song, and H.~Chen, ``Artificial intelligence in sustainable energy industry: Status quo, challenges and opportunities,'' \emph{Journal of Cleaner Production}, vol. 289, p. 125834, 2021.

\bibitem{khang2023ai}
A.~Khang, A.~Misra, S.~K. Gupta, and V.~Shah, \emph{AI-Aided IoT Technologies and Applications for Smart Business and Production}.\hskip 1em plus 0.5em minus 0.4em\relax CRC Press, 2023.

\bibitem{ravizki2023legal}
E.~N. Ravizki, L.~Yudhantaka, and R.~C.~V. Wijaya, ``Legal policy on artificial intelligent (ai): Study comparative from global practices,'' \emph{Nusantara Science and Technology Proceedings}, pp. 135--141, 2023.

\bibitem{mauritiusai}
\BIBentryALTinterwordspacing
 [Online]. Available: \url{https://ncb.govmu.org/ncb/strategicplans/MauritiusAIStrategy2018.pdf}
\BIBentrySTDinterwordspacing

\bibitem{egypt_2023}
\BIBentryALTinterwordspacing
G.~of~Egypt, ``Ministry of communications and information technology,'' 2023. [Online]. Available: \url{https://mcit.gov.eg/en/Publication/Publication_Summary/9283}
\BIBentrySTDinterwordspacing

\bibitem{rwanda_2023}
\BIBentryALTinterwordspacing
G.~of~Rwanda, ``Ministry of ict and innovation republic of rwanda,'' 2023. [Online]. Available: \url{https://www.minict.gov.rw/index.php?eID=dumpFile&t=f&f=67550&token=6195a53203e197efa47592f40ff4aaf24579640e}
\BIBentrySTDinterwordspacing

\bibitem{amankwah2022harnessing}
J.~Amankwah-Amoah and Y.~Lu, ``Harnessing ai for business development: a review of drivers and challenges in africa,'' \emph{Production Planning \& Control}, pp. 1--10, 2022.

\bibitem{lemma2022afcfta}
A.~Lemma, M.~M. Parra, and L.~Naliaka, \emph{The AfCFTA: unlocking the potential of the digital economy in Africa}.\hskip 1em plus 0.5em minus 0.4em\relax ODI, 2022, vol.~13.

\bibitem{brand2022data}
D.~Brand, J.~A. Singh, A.~G.~N. McKay, N.~Cengiz, and K.~Moodley, ``Data sharing governance in sub-saharan africa during public health emergencies: Gaps and guidance,'' \emph{South African Journal of Science}, vol. 118, no. 11-12, pp. 1--6, 2022.

\bibitem{Babalola2023}
\BIBentryALTinterwordspacing
O.~Babalola, \emph{Data Protection Legal Regime and Data Governance in Africa: An Overview"}.\hskip 1em plus 0.5em minus 0.4em\relax Springer International Publishing, 2023, pp. 71--96. [Online]. Available: \url{https://doi.org/10.1007/978-3-031-24498-8_4}
\BIBentrySTDinterwordspacing

\bibitem{shibambu2022framework}
A.~Shibambu and N.~S. Marutha, ``A framework for management of digital records on the cloud in the public sector of south africa,'' \emph{Information Discovery and Delivery}, vol.~50, no.~2, pp. 165--175, 2022.

\bibitem{kugler2022impact}
K.~Kugler, ``The impact of data localisation laws on trade in africa,'' \emph{Policy Brief}, vol.~8, 2022.

\bibitem{mukherjee2023edge}
D.~V. Mukherjee, \emph{At the Edge of Tomorrow: Unleashing Human Potential in the AI Era}.\hskip 1em plus 0.5em minus 0.4em\relax Notion Press, 2023.

\bibitem{marwala2022closing}
T.~Marwala, \emph{Closing the gap: The fourth industrial revolution in Africa}.\hskip 1em plus 0.5em minus 0.4em\relax Pan Macmillan South Africa, 2022.

\bibitem{beumer2022economic}
K.~Beumer \emph{et~al.}, ``Economic inequality and science, technology and innovation policy: The cases of the united kingdom and south africa,'' 2022.

\bibitem{roche2023ethics}
C.~Roche, P.~Wall, and D.~Lewis, ``Ethics and diversity in artificial intelligence policies, strategies and initiatives,'' \emph{AI and Ethics}, vol.~3, no.~4, pp. 1095--1115, 2023.

\bibitem{plantinga2023responsible}
P.~Plantinga, K.~Shilongo, O.~Mudongo, A.~Umubyeyi, M.~Gastrow, and G.~Razzano, ``Responsible artificial intelligence in africa: Towards policy learning,'' 2023.

\bibitem{prinsloo2022data}
P.~Prinsloo and R.~Kaliisa, ``Data privacy on the african continent: Opportunities, challenges and implications for learning analytics,'' \emph{British Journal of Educational Technology}, vol.~53, no.~4, pp. 894--913, 2022.

\end{thebibliography}
\end{document}